# New Results on Pulsating OB Stars

Paweł Moskalik

*Copernicus Astronomical Center, Polish Academy of Sciences,
ul. Bartycka 18, 00–716 Warsaw, Poland*

**Abstract.**
Until very recently the physical mechanism driving oscillations in $\beta$ Cep and other early type stars has been a mystery. The breakthrough came with the publication of new OPAL and OP opacity data. Model calculations with the new opacities have demonstrated that the pulsations are driven by the familiar $\kappa$-mechanism, acting in the metal opacity bump at $T \approx 2 \times 10^5\,$K. The mechanism excites the low order $p$- and $g$-modes in the upper part of the instability strip and the high order $g$-modes in the lower part of the strip. The theoretical instability domains agree well with the observed domains of the $\beta$ Cep and the SPB stars. In this review I present these recent theoretical results and discuss their consequences for our understanding of B stars pulsations.

## 1. Introduction

Pulsational variability is a widely spread phenomenon among O and B stars. The first group of pulsators to be identified in this part of the H–R diagram were the $\beta$ Cep stars, discovered in the begining of the century (Frost 1902; Albrecht 1908). Today 58 confirmed and 81 suspected members of this group are known (Sterken & Jerzykiewicz 1993). These early B stars display single- or multiperiodic variability of their brightness and radial velocity, with periods of $0.1 - 0.3$ day. In terms of pulsation theory such periods imply oscillations in low order $p$-modes. Several other groups of O and B pulsators have been discovered in the recent years. In 1976 Smith & Karp (see also Smith 1977) have identified the group of 53 Per stars, which show multiperiodic line profile variations with periods of the order of $1 - 2$ day. These variations are interpreted as resulting from nonradial oscillations of low spherical harmonic degree $\ell$. Photometric variability on similar time scales has been detected in several mid to late B stars by Waelkens (1991), who calls his objects the *Slowly Pulsatig B-type Stars*, or the SPB stars. Both 53 Per and the SPB stars are slow rotators and most likely they are closely related. The 53 Per itself (the only one of its group to have adequate data) is photometrically identical to the SPB stars (Smith *et al.* 1984; Huang *et al.* 1994). Line profile variations of *high* order ($\ell \geq 4$) have been discovered in $\zeta$ Oph (Walker, Yang & Fahlman 1979) and in several other early type stars. In contrast to the SPB and 53 Per-type stars, all members of this group are fast rotators. Also in this case the variability is interpreted in terms of nonradial pulsations (Voght & Penrod 1983). Another group of OB stars whose variability is believed to be due to pulsations are the hot supergiants (Maeder 1980). Finally, nonradial pulsations have also been suggested as an explanation of the periodic brightness changes observed in many Be stars (the so called $\lambda$ Eri-type stars), although the nature of this variablity is still subject of a heated debate (Baade & Balona 1994). For the detailed discussion of observational status of the OB variables the reader is referred to e.g. Balona (1995).

The origin of pulsation in the $\beta$ Cep and other OB type variables has been a long standing puzzle, challenging our undersatnding of stellar structure and stability. Already in the sixties it was realised that the opacity mechanism acting on the He$^+$ ionization zone, which is the cause of pulsation of Cepheids and RR Lyr stars (Baker & Kippenhahn 1962), cannot work in much hotter OB variables. The correct solution was first suggested by Simon (1982), who postulated that the metal opacities should be augmented by factor of $2-3$. He has pointed out that such a modification resolves the period ratio discrepancy in Cepheids, which has been another long-standing problem, and at the same time provides a driving mechanism for the $\beta$ Cep stars. Simon's suggestion has been initially met with scepticism by atomic physisists (e.g. Magee, Merts & Huebner 1984), but neverteless inspired new efforts in computing stellar opacities. As a result, two new, independently calculated, sets of opacity tables have been recently published: those of the OPAL project (Iglesias, Rogers & Wilson 1992) and those of the OP project (Seaton et al. 1994). Both sets of calculations show a large local opacity maximum at $T \approx 2 \times 10^5$K, which has been absent in the older Los Alamos tables. This new feaure is due to the combined effect of milions of metal lines and is therefore called a "metal opacity bump.

The publication of new opacities had an immediate impact on the problem of the OB star pulsations. Already the first model computations with the OPAL tables (Cox et al 1992; Moskalik & Dziembowski 1992; Kiriakidis et al. 1992) demonstrated that the excitation of the $p$-mode oscillations is possible in the $\beta$ Cep-type stellar models. These early results were not fully satifactory though. The models were unstable only in the fundamental mode (contrary to observations) and only for a somewhat high metallicity of $Z \geq 0.03$. In this early work the first version of the OPAL tables was used, in which the spin-orbit coupling in the iron atoms was not taken into account. In the subseqent release of the OPAL tables (Iglesias et al. 1992) this effect was included, which led to a further enhancemen of the metal bump. The improved tables were used in the next set of $\beta$ Cep and SPB pulsation models, computed by Dziembowski & Pamyatnykh (1993), Gautschy & Saio (1993) and Dziembowski, Moskalik & Pamyatnykh (1993). The new models, with a standard metallicity of $Z = 0.02$, show a remarkable overall agreement with the observations. The calculations performed with the OP tables give very similar results (Pamyatnykh et al. 1994). These two facts make us believe that the long sought mechanism exciting pulsations in OB star has been finally identified. In the following, I will discuss these recent theoretical results, with particular attention given to the morphology of the instability strip and to the physics of the driving. The latest review of this subject has been given by Dziembowski (1995). The new opacities have also been employed in the modeling of pulsations in the early type supergiants. In this context it is in order to mention the papers by Glatzel & Kiriakidis (1993) and by Kiriakidis et al. (1993). Unfortunaty, because of limited space, this very interesting work has to stay beyond the scope of this review.

## 2. B Stars Instability Strip

The main result of the stability surveys of stellar models built with the new opacities is the discovery of a theoretical instability strip of the B stars. In Fig. 1 we show the new instability domain in the H–R diagram. Only modes of $\ell \leq 2$, i.e. modes which are most likely to be observed *photometrically*, have been considered in determining the boundaries of the domain. The models displayed in the figure are calculated with the OP tables, assuming standard composition of $X = 0.7$ and $Z = 0.02$ (Pamyatnykh et al., in preparation). The influence of the particular choice of the opacity data (OPAL vs. OP) is discussed in Pamyatnykh et al. (1994). For comparison we also display the cluster $\beta$ Cep variables as well as the field SPB stars for which luminosities and effective temperatures are known.

The B stars instability strip begins at masses of about $3M_\odot$ (spectal type B9) and extends to masses well above $50M_\odot$, covering all range of spectral types B and O. Between this strip and the $\delta$ Sct instability strip there is only a narrow mass range, centered on $2.5M_\odot$, where the models are pulsationally stable. The $\beta$ Cep variables occupy only a small region of the strip. Nevertheless, by analogy with the *Cepheid instability strip*, we propose to call the *whole* new instability domain the *$\beta$ Cephei instability strip*.

The instability strip splits into two, partially overlaping, but otherwize distinct regions, where physically different types of oscillatory modes are driven:

### 2.1. $\beta$ Cep Domain

In this domain the instability affects the low order $p$- and $g$-modes. In a typical model we find instability of the fundamental and of the first overtone $p$-modes ($p_1$ and $p_2$) as well as of the lowest order gravity mode $g_1$. For metallicity slightly above $Z = 0.02$ the second overtone ($p_3$) might also be excited. The instability extends from $\ell = 0$ to at least $\ell = 8$. The excitation of $g_1$ mode is of particular interest, because its frequency can be a sensitive probe of the evolutionary changes in the deep stellar interior, specifically in the $\mu$-gradient zone (Dziembowski & Pamyatnykh 1991). The periods of the unstable modes range from 0.1 to 0.3 day, which is in a very good agreement with the observed period range of the $\beta$ Cep stars.

The Blue Edge of the $\beta$ Cep instability domain is very sensitive to the adopted metallicity. For $Z = 0.02$ (as in Fig. 1), the instability at lower masses $(9 - 10M_\odot)$ sets in already on ZAMS, but with the increasing mass the strip becomes narrower. This trend continues up to $20M_\odot$. For even higher mass the strip widens in both directions, finally reaching ZAMS again. Calculations of Kiriakidis et al. (1993) show that for $\log(L/L_\odot) \geq 5.8$ the $\beta$ Cep instability domain and the Cepheid instability strip merge.

The instability of the low order p-modes extends beyond the Main Sequence phase. The post-MS evolution is, however, very fast. Consequently, the observed Red Edge of the instability strip is effectively determined by the Terminal Age Main Sequence (or TAMS). Such an *Effective Red Edge* is ploted with the dashed line.

The theoretical instability strip encompasses all the observed $\beta$ Cep variables (both in clusters and in the field). The observational points do not cover the whole theoretical domain, though. There is a signifcant lack of variables with $M < 9M_\odot$. The high luminosity part of the theoretical strip also remains unpopulated, with the exception of two field stars V986 Oph and HD 34656 (Balona 1995). We do not have a good explanation for this pattern, perhaps in these parts of the H–R diagram the stars preferentially choose to pulsate in modes of high $\ell$. Such modes cannot be detected photometrically.

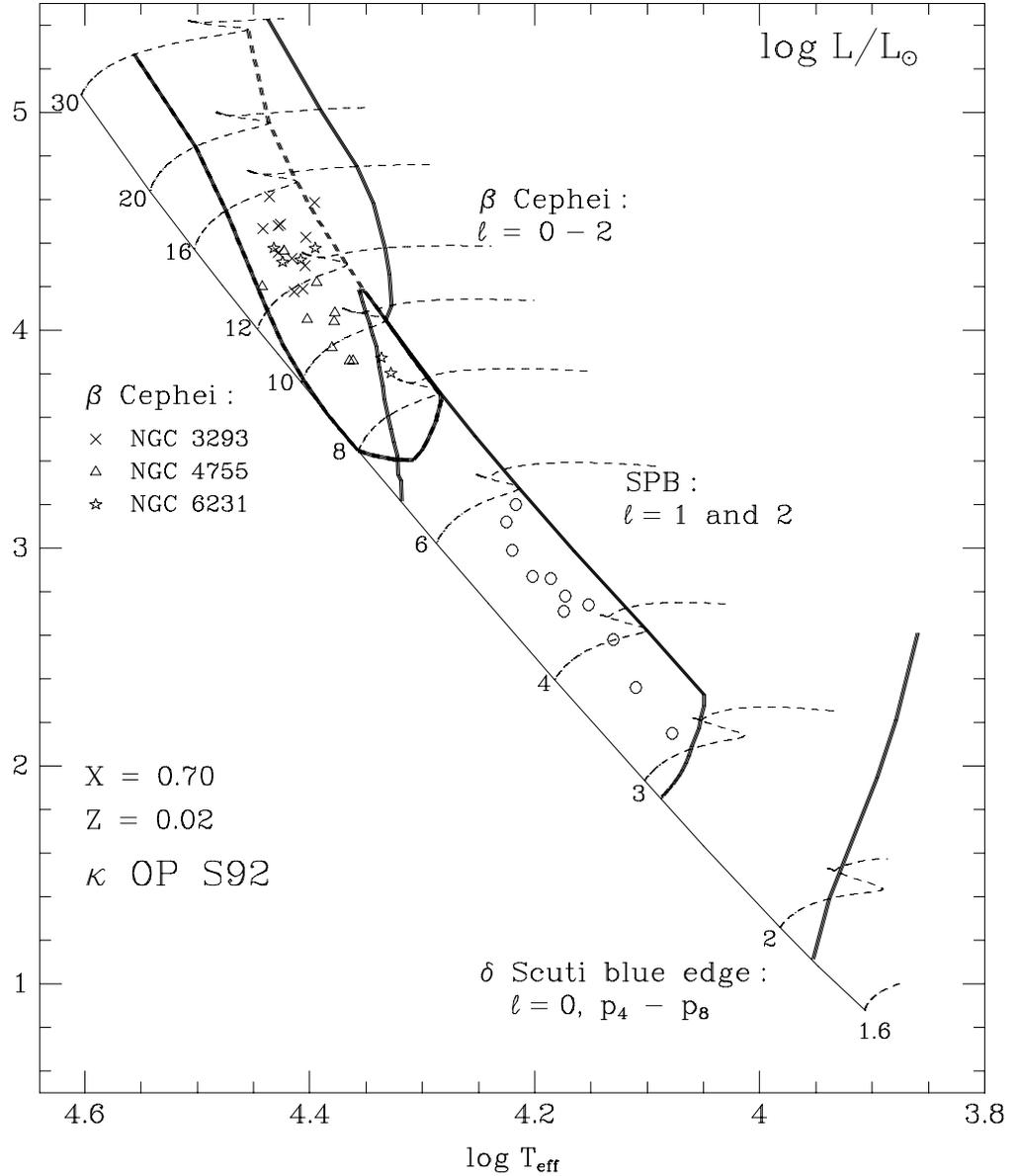

Figure 1. Theoretical instability domains of the $\beta$ Cep and the Slowly Pulsating B-type Stars in the H–R diagram. Only modes of $\ell \leq 2$ are considered. The models are computed with OP tables (S92 mixture, see Seaton et. al. 1994) for standard composition of $X = 0.7$, $Z = 0.02$. For comparison we show $\beta$ Cep stars observed in young open clusters NGC 3293, NGC 4755 and NGC 6231 (Balona 1994; Balona & Koen 1994; Heynderickx 1991) and the field SPB stars (Waelkens 1994). From Pamyatnykh et al. (in preparation).

### 2.2. SPB Domain

In this domain the instability mechanism excites the high order $g$-modes. The instability sets in at ZAMS and continues until TAMS, but not beyond. The high temperature limit of the domain (in Fig. 1 located at $T_\text{eff} \approx 4.35$) is sensitive to the metallicity and to the choice of the metal mixture. In a given model several (from 5 to 30) consecutive $g$-modes are unstable at each $\ell$. The total number and the radial order of the excited modes increases as the evolution of the star progresses. The modes are approximately equally spaced in period, as predicted by the asymptotic theory of pulsation, with the spacing depending on the mass and the age of the star. The small periodic departures from the equidistant patter are caused by partial trapping of the modes in the $\mu$-gradient zone. In that sence the SPB stars are very similar to the DOV and DBV white dwarfs (Kawaler, these proceedings) and can be studied seismologically with the same theoretical tools. The unstable modes have periods between 0.5 and 4.1 day, which agree very well with periods observed in the SPB star.

The instability is not limitted to $\ell \leq 2$. Already for $4M_\odot$ it extends to $\ell = 7$. With each increasing $\ell$ the instability domain shifts higher up the Main Sequence band, and for $\ell \geq 6$ it extends well into the $\beta$ Cep region. The excitation of these relatively high-$\ell$ modes in more massive models may be relevant to pulsations of the $\zeta$ Oph variables.

All known SPB stars fall within the theoretical instability domain displayed in Fig. 1. They populate the middle and lower parts of the domain. The 53 Per variables (not plotted) would appear in the upper part of the domain and above it. Recalling that the 53 Per stars have been detected through the line profile variations (more sensitive to higher $\ell$), while the SPB stars have been detected photometrically, it is tempting to speculate that the 53 Per stars are just the extention of the SPB phenomenon to higher $\ell$ ($\ell = 3, 4$). We would like to caution the reader, though, that with the exception of 53 Per itself, the variables of this group are not well observed. Much more observational work is needed before any final conclusion on the SPB – 53 Per connection can be drawn.

### 3. The Instability Mechanism in B Stars

Pulsational instability in B stars is driven by the familiar opacity mechanism ($\kappa$-mechanism), i.e. the same mechanism which operates in Cepheids and RR Lyr stars. However, in B stars the mechanism acts on a different opacity maximum.

For further discussion it is convinient to write down the work integral, $W$, which represents the net gain of energy by a osciallatory mode during one cycle:

$$W = -\int d^3x \, \nabla_{\rm ad} \oint dt \, \frac{\delta P}{P} \, \delta {\rm div} {\bf F_r} \qquad (1)$$

where $\delta$ denotes the Lagrangean perturbation and other symbols have their usual meaning. Eq. (1) neglects the perturbations of the nuclear reaction rate and of the convective fulux. The former effect is not important at all in B stars. The effects of convection are not very important either, since in the driving zone the convection carries very little (if any) of the total energy flux. Adopting diffusion approximation, $\delta {\rm div} {\bf F_r}$ can be expressed as

$$\delta {\rm div} {\bf F_r} = \frac{1}{4\pi r^2} \frac{d\delta L_r}{dr} \qquad (2)$$

where

$$\frac{d\delta L}{dr} = \frac{dr}{d\log T} \frac{d}{dr}\left(\frac{\delta T}{T}\right) - \frac{\delta \kappa}{\kappa} + 4\left(\frac{\delta T}{T} + \frac{\delta r}{r}\right) \qquad (3)$$

The eqs. (2) and (3) are strictly valid for radial modes and approximately valid for nonradial modes (in the later case every perturbed quantity contains angular factor $Y_l^m(\theta,\phi)$). The first term in eq. (3) describes radiative dissipation, which is the main damping effects stabilizing the modes. The $\kappa$-effect driving is due to the second term. Inserting eq. (2) and (3) into eq. (1) we obtain

$$W = \ldots + \int d^3x \frac{\nabla_{\rm ad}}{4\pi r^2} \oint dt \, \frac{\delta P}{P} \frac{d}{dr}\left(\frac{\delta \kappa}{\kappa}\right) + \ldots \qquad (4)$$

Eq. (4) shows, that the opacity perturbation contributes locally to the driving in those zones where during compression the derivative $\frac{d}{dr}(\frac{\delta \kappa}{\kappa})$ is positive. Noticing that $\delta P/P$, $\delta T/T$ and $\delta \rho/\rho$ are almost in phase and assuming that these perturbations vary slowly throughout the driving region, we find that the $\kappa$-effect *driving* occurs in places where $\kappa_T = (\frac{\partial \log \kappa}{\partial \log T})_\rho$ *increases with r*. The same effect will locally *damp* the oscillations if $\kappa_T$ *decreases with r*.

The proper behaviour of $\kappa_T$ is a necessary, but not a sufficient condition for the $\kappa$-effect to work. The other requirement is that the thermal timescale $\tau_{th}$ of the zone, defined as

$$\tau_{th} = \frac{1}{L} \int\limits_{surface}^{M(T)} T c_P dM \qquad (5)$$

is comparable or longer than the period of the mode in question. If this condition is violeted, then the zone re-adjusts quickly to the thermal equilibrium during the pulsation cycle, and consequently ${\rm div}{\bf F_r} \approx 0$.

In Fig. 2 we show two models, which are representative of the $\beta$ Cep and the SPB stars, respectivelly. All quantities are plotted as function of the temperature in the model, so that the important features in the opacities occur in the same position. In both cases the opacity patterns are similar. The two main features are the He$^+$ bump at $\log T \approx 4.6$ and the metal bump at $\log T \approx 5.3$ (the H ionization bump is absent because the surface temperature is too high). The metal bump is a new feature of the OPAL and OP opacities, which was absent in the older Los Alamos tables. In the lowest panel of the plot we show for selected modes the contribution to the work integral from different zones. The He$^+$ bump, which drives the pulsation in the classical variable stars (Cepheids, RR Lyr, $\delta$ Sct), plays no role in B stars. This is so, because the thermal timescale $\tau_{th}$ at the position of this bump is much too short. On the other hand, the timescale requirement can be satisfied at the position of the metal bump. Both $\beta$ Cep and the SPB variables are destabilized by the $\kappa$-mechanism operating on this feature.

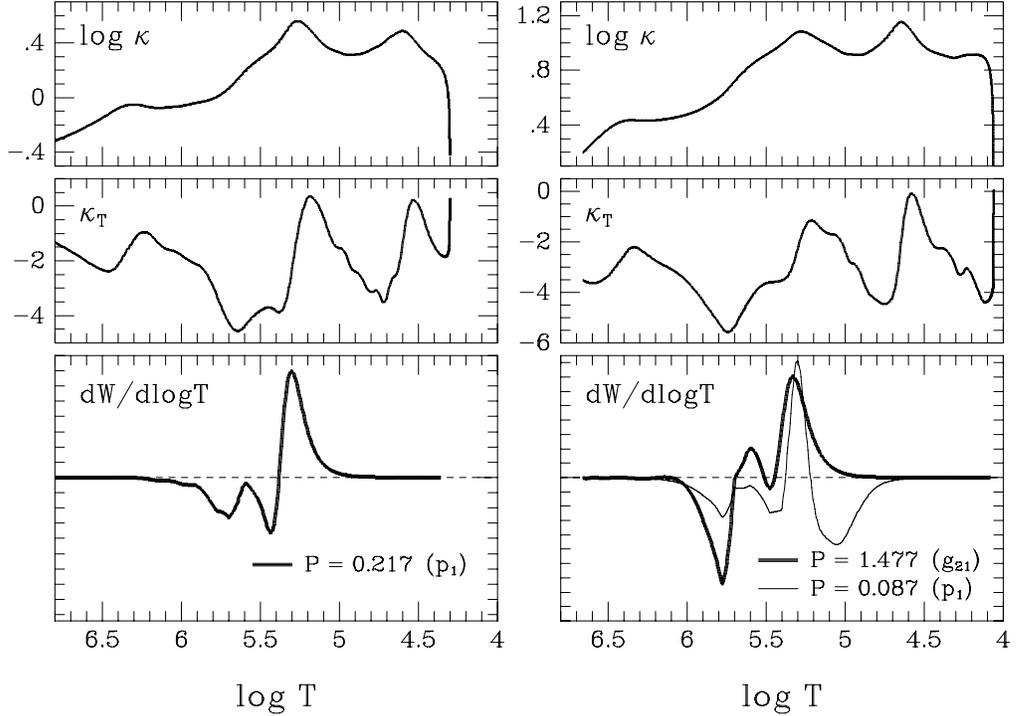

Figure 2. Opacity $\kappa$, opacity derivative $\kappa_T$, and differential work integral $dW/d\log T$ (positive in driving zone) for selected pulsation modes, plotted vs. temperature. *Left*: $\beta$ Cep star model ($M = 12 M_\odot$, $\log(L/L_\odot = 4.27$, $\log T_\mathrm{eff} = 4.378$, $X_c = 0.13$). *Right*: SPB-star model ($M = 4 M_\odot$, $\log(L/L_\odot = 2.51$, $\log T_\mathrm{eff} = 4.142$, $X_c = 0.37$). Both models have initial composition of $X = 0.7$, $Z = 0.02$. The modes plotted correspond to $\ell = 1$.

Fig. 2 explains why the same mechanism excites different types of modes in $\beta$ Cep and in the SPB stars. In the $12 M_\odot$ model $\tau_{th}$ at the zone of the metal bump is comparable to the periods of the low order $p$-modes. For these modes the local $\kappa$-effect driving is most efficient and the modes become unstable. The high order $g$-modes have considerably longer periods. Consequently, the driving at $\log T \approx 5.3$ is much less efficient and cannot overcome the radiative damping ocuring deeper in the model. Therefore, these modes in the $\beta$ Cep-type models are stable. In the case of the $4 M_\odot$ model, the thermal timescale at the metal bump is about an order of magnitude longer. The difference is caused by lower surface temperature (the zone of $\log T = 5.3$ is deeper) and lower luminosity of the model. The $\tau_{th}$ at the zone of the metal bump is now comparable to periods of the high order $g$-modes and these are the modes which become unstable. On the other hand, for the low order $p$-modes the timescale condition is satisfied already at $\log T \approx 5.0$. As a result, the zone where $d\kappa_T/dr < 0$ becomes "activated". The strong damping effect of this zone stabilizes the $p$-modes in the SPB stars, as can be seen in Fig. 2.

Finally, we need to explain why the opacity mechanism excites in the B stars either short period $p$-modes or long period $g$-modes, but never the modes with the *intermediate* periods. The thermal timescale condition at the metal bump is satisfied for such modes somewhere between $4M_\odot$ and $12M_\odot$, so why these modes are never destabilized ? The answer lies in the shape of their eigenfunction $\delta P/P$. For these modes, the eigenfunction has very small amplitude at the driving zone and considerably larger amlitude in the deep interior. As a result, the driving is too weak to overcome the radiative damping in the inner zones and the mode remains stable.

## 4.  A Few Remarks

The discovery of the metal opacity bump has provided a very simple, and fully satisfactory explanation of the B stars pulsations. The $\kappa$-mechanism operating in $\beta$ Cep and the SPB stars is fundamentally very similar to the mechanism driving pulsations of Cepheids or $\delta$ Sct stars. In contrast to the classical variables, though, our understanding of B stars pulsations is not hampered by the lack of a time dependent-theory of convection, since the convective transport plays no, or very little role in the driving region. There is a usual convection zone associated with the $He^+$ ionization zone, but this region is neutral for the driving or damping of the oscillations. This favourable situation allows us to put much more confidence in the prediction of the linear nonadiabatic pulsation theory. It also makes possible calculating reliably *all* boundaries of the B stars instability domain.

Since the driving mechanism relies on the metal opacity bump, it is quite natural that the size of the instability strip depends on the metallicity $Z$. The most affected is the Blue Edge of the strip, which becomes cooler with the decreasing $Z$. More detailed calculations show that the shape of the strip depends also on the choise of the metal mixture (Pamyatnykh et al. 1994; Pamyatnykh et al., in preparation). For the OP opacities with S92 mixture, the domain of $\beta$ Cep pulsations vanishes for $Z < 0.012$. The SPB domain is more robust and can survive at lower metallicities. It vanishes only for $Z < 0.006$. Thus, it migth be possible to see the SPB variables in the moderately metal deficient stellar populations, which are already lacking the $\beta$ Cep variables. Another consequence of the $Z$-sensitivity of the driving mechanism is that the Blue Edge of observed instability strip of the field $\beta$ Cep stars should be somewhat diffused, because of the scatter in $Z$ among the Galactic variables.

Our theory of the B star pulsations is currently limited to the linear approach. The first attempt at the nonlinear modeling has been undertaken only recently (Moskalik & Buchler 1994). Nonlinear theory is necessary to predict the amplitudes at which the instability will saturate and to understand the mode selection process. We note in passing, that in a typical $\beta$ Cep star there are 20–30 linearly unstable modes of $\ell \leq 2$ (counting the $m$-splitting), for the SPB star the respective number is well above 100. Only a small fraction of these modes reach detectable amplitudes and can be observed.

## 5. Applications

With our newly gained understanding of the B star pulsations we can use these objects as sensitive tools of stellar diagnostic. The presence of the $\beta$ Cep variables in the stellar cluster can be used to infer the metallicity of the population and to constrain the amount of the convective overshooting (Pamyatnykh et al., this conference). For an individual $\beta$ Cep star, the value of $\ell$ for each mode and the order for the radial modes can be derived from the multicolour photometry (Cugier et al. 1994). The detailed fit of the pulsation model to the observed frequency spectrum can be then performed, yielding parameters of the star. Such an attempt has been already undertaken for 16 Lac (Dziembowski & Jerzykiewicz, in preparation). Also the SPB stars, with their reach $g$-mode spectrum, can become very fruitfull targets for seismology when adequate data becomes available. The $g$-modes penetrate the whole radiative interior of the star and their frequencies are sensitive to the structure of the $\mu$-grandient zone. These stars will provide a very strong test to the convective overshooting models. The $m$-splitting of the pulsation frequencies should also enable measuring the internal rotation rate, which will constrain the theory of the angular momentum evolution.

**Acknowledgments.** This work has been supported in part by KBN (Poland) through grant No 2–P304–013–07. Generous travel grant from IAU is also acknowledged. I am greatfull to W. Dziembowski and A. Pamyatnykh for many enlightning discussions. Finally, it is my privilege to thank the organizers of the meeting for their warm hospitality during the conference.

**Discussion**

**Maeder :** In principle, as you showed, the driving of $\beta$ Cep pulsations depends very much on metallicity. Is there any evidence of differences in the amplitudes or so, related to the galactic locations of $\beta$ Cep stars ?

**Moskalik :** Pulsation amplitude is not directly connected with the growth rate (it depends on other factors as well) and cannot be used as an indicator of the strength of driving. On the other hand, with the increasing metallicity the blue edge of the instability strip becomes hotter and the $\beta$ Cep population should be on average bluer. This effect might be already observed by Waelkens *et al.* (A&A, 251, 69), who notice that $\beta$ Cep stars close to the center of the Galaxy are all fairly blue.

**Bedding :** How firm is your statement that there should be no $\beta$ Cep stars in the LMC ? Kjeldsen & Baade (IAU Symp. 162, p. 29) found tentative evidence for *low-amplitude* $\beta$ Cep pulsation in the young LMC cluster NGC 371.

**Moskalik :** As we decrease the metallicity of the models the instability strip becomes narrower and shifts progresivelly up the H–R diagram and finally disappears. The exact value of $Z$ at which the strip vanishes depends somewhat on the metal mixture and also on the choice of the opacity tables. With OP tables and S92 mixture we find $Z = 0.012$ as the treshold value. The metallicity of LMC is on average $Z \approx 0.014$ (1.4 times less than solar) and in a young cluster it might be slightly higher. According to these numbers, the $\beta$ Cep stars could be present in the LMC, but they should be very rare. On the other hand, the SMC is more metal deficient ($Z \approx 0.005$) and there should be no $\beta$ Cep stars there, unless the stellar population is locally enriched in the iron-group elements.

**M. Smith :** How do you solve the problem in your picture of avoiding a plethora of high overtone modes being excited with long periods in 53 Per (SPB) stars ?

**Moskalik :** I don't. The results, which I have presented are obtained with the linear pulsation theory, which only tells us which modes are excited. The question of mode selection and of the final pulsation amplitudes lie in the domain of a nonlinear theory. We have no credible nonlinear theory for these stars. The problem you mentioned is not specific to the SPB stars; for all the multimode pulsators ($\beta$ Cep, SPB, $\delta$ Sct, white dwarfs) we have many modes which are theoretically excited and only a small fraction of them is observed.

**M. Smith :** Suppose it were the case, as many of us in the B star community have suggested, that there are long period NR modes in late-O/early-B 53 Per stars and/or in Be stars. Would you agree in this case that an additional excitation mechanism were necessary ?

**Moskalik :** Our current theory describes nonrotating or slowly rotating stars and as such cannot be directly applied to the Be stars. Whether metal bump driving would work for these fast rotating objects is an open question. The 53 Per/SPB stars are all slow rotators, so the theory does apply here. The models predict excitation of modes with periods of $0.5 - 4.0$ day. The longer period modes cannot be driven because the thermal timescale in the zone of the metal bump is too short for them. If you observe modes with periods significantly longer than 4.0 day, then indeed an additional driving mechanism would be necessary.

**Jerzykiewicz:** This is just a comment. You cannot explain 53 Per stars by invoking high $\ell$ because they were defined by Myron Smith as $\ell = 2$ pulsators. The high $\ell$ late O and early B pulsators are called $\zeta$ Oph stars.

**Moskalik:** I had in mind the modes of $\ell$ only slightly higher, say $\ell = 3, 4$. Such modes will still cause the *low order* line profile variability, yet they will be practically undetectable photometrically, because of very strong cancellation effect.

**Bono:** What is the physical mechanism which provides the quenching of the pulsation close to the Red Edge ?

**Moskalik:** The $\beta$ Cep pulsations are driven by $\kappa$-mechanism acting on the metal opacity bump at $\log T \approx 5.3$. At lower $T_{\text{eff}}$ the metal bump is located deeper in the envelope and the thermal timescale $\tau_{th}$ of the zone becomes longer. (The periods become longer as well, but this effect is smaller.) As we approach the Red Edge of the strip, the timescale condition $\tau_{th} \approx P$ is satisfied already at $\log T < 5.3$. Consequently, the damping zone around $\log T = 5.0$ is activated and the mode becomes linearly stable. This is the same reason for which $p$-modes are stable in the SPB stars (see Fig. 2).

The answer is different for the $g$-modes in the SPB stars. In this case the Red Edge of the instability strip is given by TAMS. As the evolution progresses beyond TAMS, the star develops a wide $\mu$-gradient zone left behind the shrinking core. The high value of the Brunt-Väisälä frequency causes the $g$-mode eigenfunctions to have large amplitudes and short wavelengths here. The resulting very strong local radiative damping overcomes the envelope $\kappa$-mechanism driving.